%% file: nexpmqft-arx.tex
\documentclass[12pt]{article}
\usepackage{graphicx,epsfig,color}
\usepackage{axodraw}
\newcommand{\bc}{\begin{center}}
\newcommand{\ec}{\end{center}}
\newcommand{\be}{\begin{equation}}
\newcommand{\ee}{\end{equation}}
\newcommand{\bea}{\begin{eqnarray}}
\newcommand{\eea}{\end{eqnarray}}
\newcommand{\ba}{\begin{array}}
\newcommand{\ea}{\end{array}}
\newcommand{\lb}{\label}
\newcommand{\rf}{\ref}
\newcommand{\bfg}{\begin{figure}[htbp]}
\newcommand{\efg}{\end{figure}}
\newcommand{\pr}{Phys. Rev. }
\newcommand{\prd}{Phys. Rev. D }

\newcommand{\npb}{Nucl. Phys. B }
\newcommand{\prl}{Phys. Rev. Lett. }
\newcommand{\prp}{Phys. Rep. }

\newcommand{\cpc}{Chin. Phys. C }

\begin{document}

\vspace*{1. cm}
\bc
{\large \textbf{The $\mathbf{1/N}$ expansion method\\
\vspace{0.3 cm}
in quantum field theory}}\footnote{An introductory lecture given 
at the III$^d$ International School ``Symmetry in integrable systems
and nuclear physics'', 3-13 July 2013, Tsakhkadzor, Armenia.}\\
\vspace{1 cm}
H. Sazdjian\\
\small{Institut de Physique Nucl\'eaire, CNRS-IN2P3,\\
Universit\'e Paris-Sud, Universit\'e Paris-Saclay, 91406 Orsay, France\\
E-mail: sazdjian@ipno.in2p3.fr
}
\ec
\par
\vspace{0.75 cm}

\bc
{\large Abstract}
\ec
\par
The motivations of the $1/N$ expansion method in quantum field theory
are explained. The method is first illustrated with the $O(N)$ model 
of scalar fields. A second example is considered with the two-dimensional 
Gross-Neveu model of fermion fields with global $U(N)$ and discrete 
chiral symmetries. The case of QCD is briefly sketched.
\par  
\vspace{0.5 cm}
\par

\newpage
\setcounter{equation}{0}

\section{Introduction and motivations} \lb{s1}

Methods of resolution of quantum field theory equations are very
rare. Usually one uses perturbation theory with respect
to the coupling constant $g$, starting from free
field theory. The size of (the dimensionless) $g$ 
gives an estimate of the strength of the interaction. 
The presence in the theory of other parameters than the coupling 
constant may allow the use of perturbation theory with respect to 
those parameters. This may enlarge the possibilities of 
approximate resolution of the theory.
\par
A new parameter may emerge if the system under consideration satisfies 
symmetry properties with respect to a group of internal 
transformations. 
For example, constituents of the system (particles, nucleons, nuclei, 
energy excitations, etc.) belonging to different species, 
may have the same mass and possess the same dynamical properties 
with respect to the interaction. In such a case interchanges between
these constituents would not modify the physical properties of the
system. One might also consider transformations in infinitesimal or
continuous forms, as in the case of the rotations in ordinary space.  
The system then satisfies an invariance property under continuous 
symmetry transformations.
In many cases, the invariance might also be only approximate.
\par  
Among the continuous symmetry groups, two play an important
role in physical problems.
The first is $O(N)$, the orthogonal group,
generated by the $N\times N$ orthogonal matrices.
It has $N(N-1)/2$ parameters and generators.
The second is $SU(N)$, the unitary group, 
generated by the $N\times N$ unitary complex matrices,
with determinant equal to 1. It has $(N^2-1)$ parameters
and generators.
\par
In particle and nuclear physics, one has the approximate 
isospin symmetry group $SU(2)$ and the
approximate flavor symmetry group $SU(3)$.
Using these approximations, one can establish relations
between masses and physical parameters of various particles,
prior to solving the dynamics of the system under consideration.
\par
In the cases of the groups $O(N)$ and $SU(N)$
mentioned above, $N$ has a well defined fixed value in
each physical problem, $N=2,3,\ldots,$ etc.
It is however tempting to consider the case where $N$ is
a free parameter which can be varied at will. In particular, 
large values of $N$, with the limit $N\rightarrow\infty$,
seem to be of interest. 
At first sight, it might seem that taking large values of $N$
would lead to more complicated situations, since the number of 
parameters increases and the group representations become intricate. 
However, it has been noticed that when the limit is taken in an 
appropriate way, in conjunction with the coupling constant of the 
theory, it may lead to simpler results than the cases of finite 
$N$. If this happens, then an interesting perspective of resolution 
is opened. 
\par
One may solve the problem in the simplified situation of the limit
$N\rightarrow\infty$ and then, to improve the predictions,
consider the contributions of the terms of order $1/N$
as a perturbation.
If the true $N$ of the physical problem is sufficiently large, then 
the zeroth-order calculation done in the limit $N\rightarrow\infty$ 
would already provide the main dominant aspects of the solution of the
physical problem under consideration. This is the spirit of the 
$1/N$ expansion method in quantum field theory. We emphasize that 
generally the solution thus obtained contains nonperturbative effects
when expressed in terms of the coupling constant $g$ of the theory
and therefore it provides nontrivial insight into the dynamics of the
theory, which otherwise would be unattainable with the use of ordinary 
perturbation theory with respect to $g$.
\par
The interest of the large-$N$ limit was first noticed by Stanley 
\cite{Stanley:1968gx} in statistical physics, who used 
it in the framework of the Heisenberg model (spin-spin interactions). 
He showed that in that limit the model reduces to the spherical
approximation of the Ising model, which was soluble.
\par
The method was introduced in quantum field theory by Wilson 
\cite{Wilson:1972cf}, who applied it to the $O(N)$ model of scalar 
fields and to the $U(N)$ model of fermion fields.
\par
In 1974, 't Hooft \cite{'tHooft:1973jz,'tHooft:1974hx} applied it to 
Quantum Chromodynamics (QCD), the newly born theory of the strong 
interaction, which is a gauge theory with the non-Abelian gauge group 
$SU(3)_c$ in the internal space of color quantum numbers. (Not to be 
confused with the global flavor symmetry group $SU(3)$ met 
previously.) This theory cannot be solved with the only large-$N$ 
limit, but many simplifications occur there. In two space-time dimensions,
it is almost soluble.
\par
We shall illustrate the method by two explicit examples:
1) The $O(N)$ model of scalar fields; 
2) The two-dimensional model of Gross and Neveu with global 
$U(N)$ symmetry of fermion fields and discrete chiral 
symmetry.
\par 
Reviews and lectures on the $1/N$ expansion method can be found in Refs. 
\cite{Witten:1979kh,Witten:1979pi,Coleman:1980nk,Makeenko:1999hq}
(the list is not exhaustive).
\par

\section{The scalar $\mathbf{O(N)}$ model} \lb{s2}

The $O(N)$ model is a theory of $N$ real
scalar fields $\phi^a$ ($a=1,2,\ldots,N$),
with a quartic interaction, invariant under the
$O(N)$ group of transformations. This group is similar 
in structure to the rotation group, but acts in the internal space 
of $N$ species of the fields.
\par
The Lagrangian density is (in units where $\hbar=c=1$)
\be \lb{e1}
\mathcal{L}=\frac{1}{2}\ \partial_{\mu}\phi^a\partial^{\mu}\phi^a
-\frac{1}{2}\ \mu_0^2\phi^a\phi^a-\frac{\lambda_0}{8N}(\phi^a\phi^a)^2.
\ee
(Summation on repeated indices is understood. 
$\partial_{\mu}=\frac{\partial}{\partial x^{\mu}}$, etc.)
$\mu_0^2$ and $\lambda_0$ are real parameters,
representing the bare mass squared and the bare
coupling constant; the physical mass and coupling constant could
be defined only after quantum radiative corrections are taken into
account. In four space-time dimensions the coupling constant
is dimensionless. The factor $1/N$ has been explicitly 
introduced in the interaction term for future convenience.
When the limit $N\rightarrow\infty$ is taken, the coupling constant
$\lambda_0$ will be assumed to be independent of $N$. Had we defined
a coupling constant $\tilde{\lambda}_0$ as being equal to
$\lambda_0/N$, we would be obliged later, to maintain a physical
content for the theory, to assume that the product $N\tilde{\lambda}_0$
remains finite in the above limit, which leads back to our initial choice.
\par
A detailed study of this model can be found in Ref. \cite{Coleman:1974jh}.
\par 

\subsection{Classical approximation} \lb{s21}

We search for the ground state of the theory at the classical level.
The lagrangian density is composed of the kinetic energy density minus
the potential energy density. The kinetic energy term gives 
generally a positive contribution to the total energy of the system.
Its minimum value is zero, corresponding to constant fields. We stick
to that situation. The potential energy density is then
\be \lb{e2}
U=\frac{1}{2}\ \mu_0^2\phi^a\phi^a+\frac{\lambda_0}{8N}(\phi^a\phi^a)^2.
\ee
This is a quartic function of the $\phi$s. It is evident that the 
energy is not bounded from below if $\lambda_0<0$. In this
case there is no a stable ground state.
We assume henceforth that $\lambda_0>0$.
\par
Two cases have to be distinguished: 1) $\mu^2_0>0$;
2) $\mu^2_0<0$.
\par
1) $\mu^2_0>0$.
\par
The shape of the function $U$ is presented in Fig. \rf{f1}.
\par
\bfg
\bc
\epsfig{file=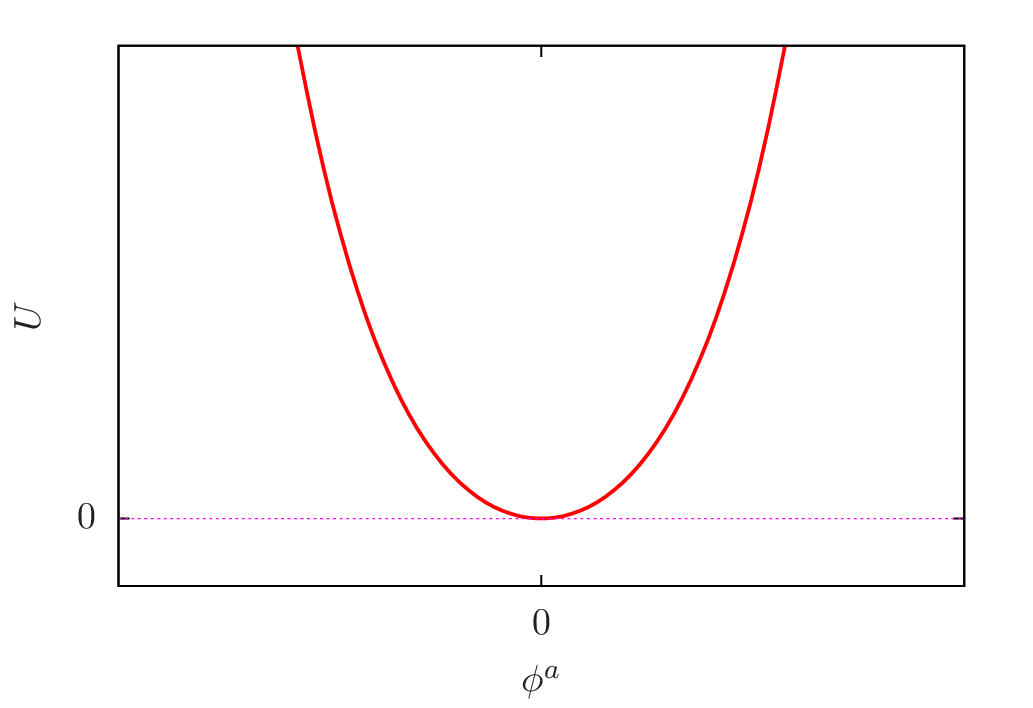,scale=1.}
\caption{The potential energy density for $\mu^2_0>0$.}
\lb{f1}
\ec
\efg
The minimum of the energy corresponds to the values
$\phi_0^a=0,\ (a=1,\ldots,N)$. The $\phi$s
in this case can be considered as excitations from the ground state.
\par
Considering the lagrangian density (\rf{e1}), we can interpret it as 
describing the dynamics of $N$ particles with degenerate masses,
equal to $\mu_0$, interacting by means of the quartic
interaction. We have complete $O(N)$ symmetry between
the particles.
\par
This mode of symmetry realization is called the Wigner mode
or the normal mode.
\par
2) $\mu^2_0<0$.
\par
The shape of the function $U$ is presented in Fig. \rf{f2}. 
\par
\bfg
\bc
\epsfig{file=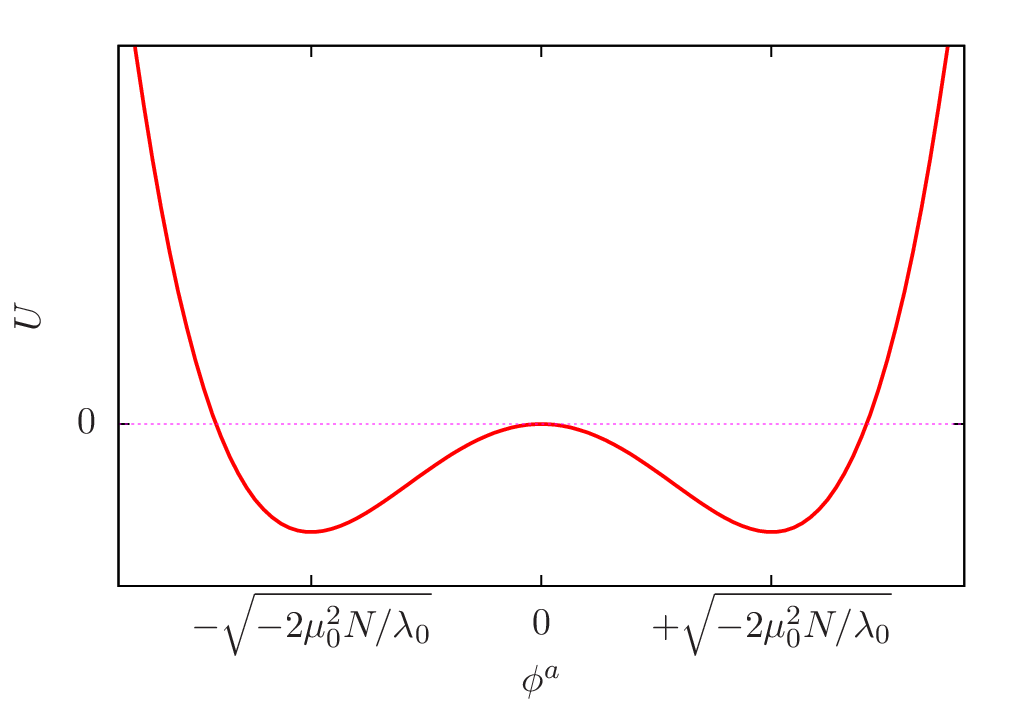,scale=1.2}
\caption{The potential energy density for $\mu^2_0<0$.}
\lb{f2}
\ec
\efg
The minimum of the energy is no longer $\phi_0^a=0$, 
but a shifted value at 
$\phi_0^a=\pm\sqrt{-\frac{2\mu_0^2N}{\lambda_0}}\equiv <\phi>$ 
for one of the $a$\,s. There is a degeneracy between the
different $\phi^a$s. For definiteness, we choose the ground state
at the minimum obtained with $\phi^N$.
\par
We redefine the fields around the new ground state:
\be \lb{e3}
\chi=\phi^N-<\phi>,\ \ \ \ \ \ \ \pi^a=\phi^a,\ \ \ a=1,\ldots,N-1.
\ee 
The potential energy density becomes
\be \lb{e4}
U=-\mu_0^2\chi^2+\frac{\lambda_0}{2N}<\phi>\chi^3
+\frac{\lambda_0}{8N}(\pi^a\pi^a+\chi^2)^2.
\ee
The fields $\pi^a$ have no longer mass terms, while
the field $\chi$ has a mass term. We have now the
masses:
\be \lb{e5}
m_{\pi^a}^2=0,\ \ \ \ a=1,\ldots,N-1,\ \ \ \ \ \ m_{\chi}^2=-2\mu_0^2.
\ee
\par
The $O(N)$ symmetry that we had initially has 
partially disappeared. There is now $O(N-1)$ symmetry
in the space of the fields $\pi^a$.
It is said that the $O(N)$ symmetry has been
spontaneously broken. This happened because the ground state
of the theory is not symmetric, while the lagrangian density is.  
\par
This phenomenon is accompanied with the appearance of $(N-1)$ massless
fields. These are called Goldstone bosons. 
This way of realization of the symmetry is called the Goldstone
mode.
\par
In particle and nuclear physics, the isospin $SU(2)$
symmetry and the quark flavor $SU(3)$ symmetry are
realized with the Wigner mode.
The chiral $SU(3)_R\times SU(3)_L$ symmetry is realized
with the Goldstone mode.
\par  

\subsection{Quantum effects} \lb{s22}

We want now to take into account the quantum effects of the model
that we are considering. For this, it is necessary to compute the
quantum corrections that contribute to the potential energy.
In  quantum field theory, these are represented by the radiative
corrections.
\par
The definition of the potential energy density $U$ is enlarged. 
The new potential energy is called the effective potential, which
is composed of the classical part, $U_{\mathrm{class}}$, that we met 
before [Eq. (\rf{e2})], and of a part, $U_{\mathrm{rad}}$,
coming from the radiative corrections:
\be \lb{e6}
U_{\mathrm{eff}}=U_{\mathrm{class}}+U_{\mathrm{rad}}\ .
\ee
$U_{\mathrm{rad}}$ is best defined in the path integral formalism.
We refer the reader to Ref. \cite{Abers:1973qs} and to the many 
textbooks that exist on the subject.
The key object is the generating functional of one-particle irreducible
diagrams or proper vertices. The effective potential is obtained 
from the latter by considering only external constant fields in
$x$-space or external lines with zero momenta in momentum space. 
Diagramatically, $U_{\mathrm{rad}}$ is given by the sum of all loop 
diagrams with such external lines. These are also accompanied with 
appropriate combinatorial factors due to the existing symmetry properties
under exchanges among the external lines \cite{Coleman:1973jx}. 
Another method of calculation hinges on a direct evaluation of the 
path integral contribution, avoiding explicit summation of diagrams, 
\cite{Jackiw:1974cv}. However, whatever the method of evaluation is,
the exact calculation of the effective potential is almost impossible,
since it involves an infinity of many complicated contributions.
Nevertheless, the use of the $1/N$ expansion method considerably  
simplifies the situation. The leading terms of this
expansion are calculable.   
\par
Prior to the evaluation of the effective potential, we shall 
introduce the propagators, vertices and loops, that are needed
for our calculations.
\par 

\subsection{Propagators, vertices and loops} \lb{s23}

The radiative corrections can be calculated starting from the
situation where $\mu_0^2>0$. Radiative corrections 
will modify the value of $\mu_0^2$, bringing it into 
a new value $\mu^2$. It is then sufficient to consider
at the end the analytic continuation of $\mu^2$ into 
negative values to complete the study.
We therefore consider the initial Lagrangian density (\rf{e1}):
\be \lb{e7}
\mathcal{L}=\frac{1}{2}\ \partial_{\mu}\phi^a\partial^{\mu}\phi^a
-\frac{1}{2}\ \mu_0^2\phi^a\phi^a-\frac{\lambda_0}{8N}(\phi^a\phi^a)^2.
\ee
\par
The inverse of the free propagator of the field $\phi^a$ 
is essentially represented by the coefficients of the quadratic 
parts of $\mathcal{L}$. In momentum space, the free propagator is  
\be \lb{e8}
D_0^{ab}(p)=\delta_{ab}D_0(p)=\int d^4x e^{{\displaystyle ip.x}}
<0|T(\phi^a(x)\phi^b(0))|0>,
\ee
where the last term represents the vacuum expectation value of
the chronological product of the field operators, and $D_0$ 
has the expression
\be \lb{e9}
D_0(p)=\frac{i}{p^2-\mu_0^2+i\varepsilon}
\hspace{1 cm}
\Longleftrightarrow
\hspace{1 cm}
\input{f3-nexpm.pstex_t}
\ \ \ \ \ \ .
\ee
We have associated with the propagator a graphical representation
in the form of a full straight line.
\par
The bare vertex is equal to the coefficient of the four-field
interaction term (contact interaction), with a multiplicative  
factor $i$ (Fig. \rf{f3}). Its order in $N$, for large values 
of $N$, is $N^{-1}$.
\par
\bfg
\bc
\input{f4-nexpm.pstex_t}
\caption{The bare vertex. We have explicitly indicated its order
in $N$.}
\lb{f3} 
\ec
\efg
Radiative corrections are represented by loops, made of closed 
lines (propagators). Figs. \rf{f4} and \rf{f5} represent examples
of one-loop and two-loop diagrams, respectively. The order in $N$
of a loop diagram is calculated by taking into account that of the
vertex at the contact point of the loop with the external lines and
that of the possibly existing summation of indices of the loop. Thus,
if the index $b$ of the propagators of a loop is independent of the
index $a$ of external lines, then it is summed over the $N$ values
the index $b$ can take; $b$ in this case is a dummy index; therefore 
it produces a multiplicative factor $N$. External lines are not
counted. 
\par
\bfg
\bc
\input{f5-nexpm.pstex_t}
\caption{Examples of one-loop diagrams with their order in $N$.}
\lb{f4}
\ec
\efg
\par
\bfg
\bc
\input{f6-nexpm.pstex_t}
\caption{Examples of two-loop diagrams.}
\lb{f5}
\ec
\efg
\par

\subsection{The auxiliary field method} \lb{s24}

We observe on Fig. \rf{f4} that the last two diagrams, which have
the same topological structure, have different behaviors for large
$N$, depending on the values of indices the loop propagators have.
This is an annoying situation, because at higher orders (many loops)
it becomes more and more difficult to continue the analysis, since
the number of possibilities the indices can have rapidly increases 
with a corresponding increase of different categories of behavior in 
$N$. The ideal situation is the one in which all diagrams with the 
same topological structure have the same behavior in $N$; in such a 
case, we do not need to consider the detailed values of indices of 
each line. 
\par
To remedy the present difficulty, we shall resort to a
method, called the auxiliary field method, often used in quantum field 
theory, which consists of replacing composite fields by a new, 
nonpropagating field, without changing the physical content of the 
theory. The validity of the latter property is rather easily shown
in the path integral formalism; when an operator formalism is used,
a simple hint is provided with the use of the equations of motion.
\par
We wish to replace the composite field $\phi^a\phi^a$ appearing in
the interaction term by a single field, which we shall designate by 
$\sigma$. (Since in $\phi^a\phi^a$, $a$ is summed from one to $N$,
this field does not have any index; it is a singlet under the $O(N)$
group of transformations.) To this end, we add to the Lagrangian density 
(\rf{e7}) a new term, thus defining a new Lagrangian density: 
\be \lb{e10}
\mathcal{L'}=\mathcal{L}+\frac{N}{2\lambda_0}\ 
\Big(\sigma-\frac{\lambda_0}{2N}\ \phi^a\phi^a-\mu_0^2\Big)^2.
\ee
The $\sigma$ field does not have kinetic energy. The terms in the
added expression are chosen such that the quartic term in $\phi$ 
as well as the mass term of $\phi$ disappear from the new Lagrangian 
density. The equation of motion (Euler-Lagrange equation) of 
$\sigma$ yields its definition:
\be \lb{e11}
\sigma=\frac{\lambda_0}{2N}\ \phi^a\phi^a+\mu_0^2.
\ee
The latter also shows that the added term in the lagrangian is zero
if the equation of motion of $\sigma$ is used. One may conclude that  
the new lagrangian is equivalent to the former one:
\be  \lb{e12}
\mathcal{L'}\approx \mathcal{L}. 
\ee
$\mathcal{L'}$ takes the form:
\be \lb{e13}
\mathcal{L'}=\frac{1}{2}\ \partial_{\mu}\phi^a\partial^{\mu}\phi^a
+\frac{N}{2\lambda_0}\sigma^2-\frac{1}{2}\sigma\phi^a\phi^a-
\frac{N\mu_0^2}{\lambda_0}\sigma.
\ee
We have now two types of field, the $\phi$s and $\sigma$, with an
interaction term between them which is no longer quartic. 
\par
We reanalyze the properties of the propagators, the vertex and the
loops with the new Lagrangian.
\par
The bare propagators are
\be \lb{e14}
D_{0\phi}(p)=\frac{i}{p^2+i\varepsilon}\ \ \Longleftrightarrow \ \ 
\input{f3-nexpm.pstex_t}\ \ \ = O(N^0),
\ee
\be \lb{e15}
D_{0\sigma}(p)=\frac{i\lambda_0}{N}\ \ \Longleftrightarrow \ \ 
\input{f7-nexpm.pstex_t}\ \ \ = O(N^{-1}).
\ee
\par
The bare vertex $\sigma\phi\phi$, with coefficient $-i/2$, is 
represented in Fig. \rf{f6}.
\par
\bfg
\bc
\input{f8-nexpm.pstex_t}
\caption{The vertex $\sigma\phi\phi$ and its order in $N$.}
\lb{f6}
\ec
\efg
Loops and higher-order diagrams are represented in Figs. \rf{f7}
and \rf{f8}.
\par
\bfg
\bc
\input{f9-nexpm.pstex_t}
\caption{Higher-order diagrams and their order in $N$.}
\lb{f7}
\ec
\par
\efg
\bfg
\bc
\input{f12-nexpm.pstex_t}
\caption{Higher-order diagrams and their order in $N$.}
\lb{f8}
\ec
\efg
\par
In the analysis of the above diagrams external lines are not counted.
We observe now many new properties with respect to the former situation
of Sec. \rf{s23}. 
First, because of the behavior of the $\sigma$ propagator as $O(N^{-1})$,
every internal $\sigma$ line introduces an additional factor $N^{-1}$
in the behavior of the corresponding diagram. Therefore, diagrams
containing an increasing number of internal $\sigma$ lines become 
nondominant. The leading diagrams in their behavior in $N$ will be those 
that contain the least possible number of internal $\sigma$ lines.
Second, loops of $\phi$s can occur only if they are
joined to $\sigma$ lines or propagators or within the latter. Multiloop
diagrams of $\phi$s are of the type of the first diagram of Fig. \rf{f8}
and of its generalizations. However, such diagrams, as well as the
second diagram of Fig. \rf{f7}, are not of the one-particle irreducible 
type, since by cutting one internal $\sigma$ propagator, the diagrams 
become separated into two disconnected diagrams. Therefore,
they cannot enter into the definition of the effective
potential. The only loop diagrams of $\phi$s to be considered are 
represented by the third diagram of Fig. \rf{f7} and the last two 
diagrams and their generalizations of Fig. \rf{f8}. The latter, 
containing internal $\sigma$ lines, are nondominant with respect to
the first one.   
\par
In conclusion, in the limit of large $N$, the leading loop 
contributions to the effective potential will come from the third 
type of diagram of Fig. \rf{f7} and of its generalizations, where 
one can have any number of external $\sigma$ lines (cf. Fig. \rf{f9}).
\par

\subsection{Renormalization} \lb{s25}

The effective potential can now be calculated. As expressed in Eq.
(\rf{e6}), it is composed of two parts, $U_{\mathrm{class}}$ and
$U_{\mathrm{rad}}$, corresponding to the classical part and to the 
part receiving contributions of radiative corrections, respectively:
\be \lb{e16}
U_{\mathrm{eff}}=U_{\mathrm{class}}+U_{\mathrm{rad}}\ .
\ee
\par
The classical part is fixed by the content of the Lagrangian density
(\rf{e13}):
\be \lb{e17}
U_{\mathrm{class}}=-\frac{N}{2\lambda_0}\sigma^2+
\frac{1}{2}\sigma\phi^a\phi^a+\frac{N\mu_0^2}{\lambda_0}\sigma.
\ee
\par
The radiative corrections are given by the sum of $\phi$ loop 
diagrams with an increasing number of external $\sigma$ lines
(Fig. \rf{f9}). Here, according to the definition of the effective
potential, the external $\sigma$ fields should be considered as 
constants in $x$-space, or, equivalently, carrying zero momenta in
momentum space.
\par 
\bfg
\bc
\input{f10-nexpm.pstex_t}
\caption{Diagrams contributing to the radiative corrections in the
effective potential at leading order in $N$.}
\lb{f9} 
\ec
\efg
\par
The summation of the above diagrams, with appropriate combinatorial
factors, can be done with conventional methods 
\cite{Coleman:1974jh,Coleman:1973jx}. The loop calculation involves
a four-dimensional integration in Minkowski space of $\phi$ propagators
(\rf{e14}). Because of the presence of the additive factor 
$i\varepsilon$ in the denominator of the propagator, the latter has a 
well-defined analyticity property, which allows one to rotate the
$k_0$-integration from the real axis to the imaginary axis and to 
calculate the integrals in Euclidean space. This amounts to replacing 
in the integrals $k_0$ with $ik_4$, with $k_4$ real. In the following,
we shall write the integrals directly in Euclidean space, with the
definition $k^2=\sum_{i=1}^4k_i^2>0$.
\par
After summation of the diagrams, the effective potential takes the
form
\be \lb{e18}
U_{\mathrm{eff}}=-\frac{N}{2\lambda_0}\sigma^2
+\frac{1}{2}\sigma\phi^a\phi^a + \frac{N\mu_0^2}{\lambda_0}\sigma + 
\frac{N}{2}\int \frac{d^4k}{(2\pi)^4}\ln(k^2+\sigma).
\ee
\par
We are interested by the stationary point of $U_{\mathrm{eff}}$
(ground state). We therefore calculate its partial derivatives
with respect to $\sigma$ and $\phi^a$:
\be \lb{e19}
\frac{\partial U_{\mathrm{eff}}}{\partial\sigma}=
-\frac{N}{\lambda_0}\sigma+\frac{1}{2}\phi^a\phi^a
+\frac{N\mu_0^2}{\lambda_0} + \frac{N}{2}\int \frac{d^4k}{(2\pi)^4}
\frac{1}{(k^2+\sigma)}\ ,
\ee
\be \lb{e20}
\frac{\partial U_{\mathrm{eff}}}{\partial\phi^a}=\phi^a\sigma,\ \ \ \ 
\ \ \ a=1,\ldots,N.
\ee
\par
The integral that appears in Eq. (\rf{e19}) is divergent for values of 
$k\rightarrow\infty$. This is a general problem of quantum field 
theory, which reflects the singular behavior of the theory at high
energies or at short distances between the fields in $x$-space.
To cure this difficulty, one generally tries to absorb the divergent 
parts into the bare coupling constant and the mass, as well as into 
redefinitions of the fields, defining finite quantities. If this 
happens, then the theory is classified as being of the renormalizable 
type. However, not all theories satisfy this requirement. There might 
arise divergences, with specific structure, which could not be absorbed 
by existing quantities. As we shall see, the present theory is of the
renormalizable type.  
\par 
To study the possible renormalizability of the theory, we isolate in 
the integrand of the above integral the dominant parts of the asymptotic 
behavior:
\be \lb{e21}
\frac{1}{k^2+\sigma}_{\stackrel{{\displaystyle \rightarrow}}
{k^2\rightarrow\infty}}\ \frac{1}{k^2}-\frac{\sigma}{(k^2)^2}
+O(1/(k^2)^3).
\ee
The first two terms lead to ultraviolet divergences by integration.
We cannot, however, manipulate them as they stand, since the second term
would lead to a new artificial infrared divergence (when $k\rightarrow 0$).
We have to incorporate in the second term a mass factor in the 
denominator to render it softer in the infrared region. 
\par
We add and subtract in 
$\frac{\partial U_{\mathrm{eff}}}{\partial\sigma}$ the following 
quantities:
\be \lb{e22}
\frac{N}{2}\int \frac{d^4k}{(2\pi)^4}\frac{1}{k^2}
-\frac{N}{2}\sigma\int \frac{d^4k}{(2\pi)^4}\frac{1}{k^2(k^2+M^2)},
\ee
where $M$ is an arbitrary mass term.
\par
The result is
\bea \lb{e23}
\frac{\partial U_{eff}}{\partial\sigma}&=&
-N\sigma\Big(\ \frac{1}{\lambda_0}+\frac{1}{2}
\int \frac{d^4k}{(2\pi)^4}\frac{1}{k^2(k^2+M^2)}\ \Big)
+\frac{1}{2}\phi^a\phi^a\nonumber \\
& &\ +N\Big(\ \frac{\mu_0^2}{\lambda_0}
+\int \frac{d^4k}{(2\pi)^4}\frac{1}{k^2}\ \Big)
+\frac{N}{32\pi^2}\sigma\ln(\frac{\sigma}{M^2}).
\eea
The infinite integrals, together with the bare coupling constant and 
the bare mass term, may define finite renormalized quantities. This
would be possible if we admit that the bare quantities $\lambda_0$
and $\mu_0^2$ may eventually take unphysical values. We thus define 
a finite coupling constant $\lambda$ and a finite mass term $\mu^2$
with the equations
\be \lb{e24}
\frac{1}{\lambda(M)}=\frac{1}{\lambda_0}+\frac{1}{2}
\int \frac{d^4k}{(2\pi)^4}\frac{1}{k^2(k^2+M^2)},
\ee
\be \lb{e25}
\frac{\mu^2(M)}{\lambda(M)}=\frac{\mu_0^2}{\lambda_0}
+\int \frac{d^4k}{(2\pi)^4}\frac{1}{k^2}.
\ee
\par
The finite quantities depend on the arbitrary mass parameter 
$M$. For each fixed value of $M$, they take a corresponding value. 
\par
One can better study this dependence by comparing for instance the
value of the coupling constant $\lambda(M)$ to a
reference value $\lambda_1\equiv\lambda(M_1)$ corresponding 
to a reference mass parameter $M_1$. 
One obtains from the comparison of the corresponding two equations:
\be \lb{e26}
\frac{1}{\lambda(M)}=\frac{1}{\lambda_1}-\frac{1}{32\pi^2}
\ln(\frac{M^2}{M_1^2}).
\ee
This equation is called the renormalization group equation (RGE)
and plays an important role for the analysis and understanding of the
properties of the theory.
\par
We can also express $\lambda(M)$ in terms of $\lambda_1$ and the other
parameters: 
\be \lb{e27}
\lambda(M)=\frac{\lambda_1}{1-\frac{\lambda_1}{32\pi^2}
\ln(\frac{M^2}{M_1^2})}.
\ee
$\lambda_1$ and generally $\lambda(M)$ are assumed positive for 
physical reasons (below boundedness of the energy; see the discussion
of the classical potential energy density after Eq. (\rf{e2})).
Therefore, we consider domains of solution where this condition is
satisfied.  
\par
From the last equation we deduce that when $M$ increases starting 
from $M_1$, $\lambda(M)$ increases. When $M$ reaches
the value $M_{\mathrm{cr}}=M_1\exp(16\pi^2/\lambda_1)$,   
$\lambda(M)$ diverges and for larger values of $M$, $\lambda$ becomes
negative. This is a sign of the instability
of the theory for large values of $\lambda$.
\par
The RGE can also be formulated in the form of a differential equation.
Defining
\be \lb{e28}
M\frac{\partial \lambda(M)}{\partial M}=\beta(\lambda(M)),
\ee
one finds from the finite form of $\lambda(M)$ [Eq. (\rf{e27})]
\be \lb{e29}
\beta(\lambda)=\frac{\lambda^2}{16\pi^2}>0,
\ee
which shows that $\lambda(M)$ is an increasing function
of $M$.
\par
What is the interest of the RGE?
In perturbation theory, when calculating radiative corrections, one usually
finds powers of the quantity $\lambda^2\ln(p^2/M^2)$,
where $p$ is a representative of the momenta of the
external particles. Even if $\lambda$ is chosen small,
at high energies, i.e., at large values of $p$,
the logarithm may become large enough to invalidate perturbative 
calculations.  
However, since the logarithm depends on $M$ for 
dimensional reasons, one can choose the latter of the order of 
$p$ to maintain the logarithm small. This could be useful, 
if on the other hand $\lambda$ remains bounded or small 
at these values of $M$. The RGE precisely gives us the 
answer to that question, indicating the way $\lambda$ 
behaves at large values of $M$.
\par
Coming back to our theory, we see from the previous results that this
is not the case here: $\lambda$, on the contrary, 
increases with $M$ and diverges at $M_{\mathrm{cr}}$.
The theory is ill-defined at high energies. In $x$-space,
high energies correspond to short distances. The interaction becomes 
stronger when the distance between two sources or two particles 
decreases. A similar behavior is also found in Quantum Electrodynamics
(QED). These theories are better defined at low energies or large
distances.
\par
Theories for which the coupling constant decreases at high energies
or at short distances are called asymptotically free.
\par 
Summarizing the results of this section, we recall that we could get 
rid of the divergent parts of the integrals by absorbing them in the bare
coupling constant and the bare mass term, redefining at the
end a finite coupling constant and a finite mass term. The theory
we are considering is therefore renormalizable. The price that is paid 
is the introduction of an arbitrary mass parameter that essentially
fixes the mass scale of the physical quantities of the theory.  
\par 

\subsection{The ground state} \lb{s26}

Using Eqs. (\rf{e24}) and (\rf{e25}), one can express the effective
potential and its derivatives [Eqs. (\rf{e18}), (\rf{e19}) and 
(\rf{e20})] in terms of the finite coupling constant and mass term.
To simplify the notation, we denote $\lambda(M)=\lambda$ and
$\mu^2(M)=\mu^2$.
\par
The effective potential is
\be \lb{e30}
U_{\mathrm{eff}}=\frac{1}{2}\phi^a\phi^a\sigma
-\frac{N\sigma^2}{2\lambda}(1+\frac{\lambda}{64\pi^2})
+\frac{N\mu^2\sigma}{\lambda}
+\frac{N\sigma^2}{64\pi^2}\ln(\frac{\sigma}{M^2}).
\ee
\par
The equations defining the minimum of the effective potential are
\be \lb{e31}
\frac{\partial U_{\mathrm{eff}}}{\partial\sigma}=
-\frac{N\sigma_0}{\lambda}+\frac{1}{2}\phi_0^a\phi_0^a
+\frac{N\mu^2}{\lambda}
+\frac{N}{32\pi^2}\sigma_0\ln(\frac{\sigma_0}{M^2})=0,
\ee
\be \lb{e32}
\frac{\partial U_{eff}}{\partial\phi^a}=\phi_0^a\sigma_0=0,\ \ \ \ \ \ 
a=1,\ldots,N.
\ee
\par
From Eqs. (\rf{e32}), one deduces two possibilities: 
1) $\phi_0^a=0$; 2) $\sigma_0=0$.
\par
1) $\phi_0^a=0$\ \ ($a=1,\ldots,N$), $\sigma_0\neq 0$.
From Eq. (\rf{e31}) one obtains
\be \lb{e33}
\sigma_0\Big(1-\frac{\lambda}{32\pi^2}\ln(\frac{\sigma_0}{M^2})\Big)
=\mu^2.
\ee
\par
The presence of the logarithm imposes $\sigma_0>0$.
For small values of $\lambda$, one has 
$\mu^2>0$. By continuity to larger values, one should
search for positive solutions of $\sigma_0$.
Let $\sigma_0$ be the solution of the equation for
a certain domain of $\lambda$ and $\mu^2$.
One should redefine the field $\sigma$ from the value of 
$\sigma_0$:
\be \lb{e34}
\sigma(x)=\sigma'(x)+\sigma_0.
\ee  
The corresponding shift of $\sigma$ in the Lagrangian
gives back a common mass to the fields $\phi^a$. 
We are in a situation where the $O(N)$ symmetry is
realized in the Wigner mode.
\par 
2) $\sigma_0=0$.
From Eq. (\rf{e31}) one obtains
\be \lb{e35}
\frac{1}{2}\phi_0^a\phi_0^a+\frac{N\mu^2}{\lambda}=0.\ \ \ \
\Longrightarrow\ \ \ \ \ \mu^2<0.
\ee
\be \lb{e36}
\phi_0^a\phi_0^a=-\frac{2N\mu^2}{\lambda}.
\ee
There is a degeneracy of solutions for the $\phi_0^a$s.
One can choose for example:
\be \lb{e37}
\phi_0^a=0,\ \ \ \ a=1,\ldots,N-1,\ \ \ \ \ \ 
\phi_0^N=\sqrt{-\frac{2N\mu^2}{\lambda}}\equiv <\phi>.
\ee
One then develops $\phi^N$ around $<\phi>$:
\be \lb{e38}
\phi^N(x)=\chi(x)+<\phi>.
\ee
\par
The symmetry is realized in the Goldstone mode, with the presence
of $(N-1)$ massless fields $\phi^a$ and one massive field $\chi$.
\par
We summarize the results obtained so far.
The $O(N)$ model defines a renormalizable theory. 
The radiative corrections introduce limitations into the validity 
domain of the model.
The coupling constant increases at high energies and diverges at some
critical mass scale.
The ground state equations also introduce new constraints on the
parameters of the model.
The two modes of realization of the symmetry, found at the classical
level, remain valid within the restricted domain of the parameters.
The above  results are obtained at the leading-order of the 
$1/N$ expansion, which allows us to simplify in a consistent way 
the equations of the theory, to solve them and to have an insight on 
the dynamics, going beyond ordinary perturbation theory. Unfortunately,
the $O(N)$ model is not a stable theory as a whole, and the continuation
of the investigations at nonleading orders of the expansion in $1/N$ 
reveals new restrictions and instabilities. 
\par

\section{The Gross-Neveu model} \lb{s3}

\subsection{General properties} \lb{s31}

We consider now an analog of the $O(N)$ model with the boson fields
replaced by fermion fields of $N$ different species with the same mass.
Since Dirac fermion fields are not generally hermitian, the symmetry
group of transformations that leaves the Lagrangian density invariant
is $U(N)$.
Theories with four-fermion interactions are not 
renormalizable in four space-time dimensions. This is related
to the fact that the fermion fields have mass dimension 3/2, instead of
1 for the boson fields. The interaction term has thus dimension
6, greater than the dimension of the Lagrangian density which is 4 
(dimensions of multiplicative coefficients are
not considered here). Such terms do not lead to renormalization.
Historically, a model of this kind was considered by Nambu and
Jona-Lasinio \cite{Nambu:1961tp} to implement dynamical chiral symmetry 
breaking. This model, because of its simplicity and of its ability
to describe nontrivial dynamical phenomena, is used until now as a 
guiding tool by many authors; on the other hand, because of its 
nonre\-nor\-ma\-li\-za\-bi\-li\-ty, it cannot represent a consistent 
theory, unless it is embedded in a wider
theory from which it would emerge as an approximation.
\par      
The renormalization difficulty can, however, be circumvented by
considering the model in two space-time dimensions. Here, the
fermion fields have dimension 1/2 and the interaction term has
dimension 2, equal to the dimension of the Lagrangian density.
This ensures the renormalizability of the theory. On the other hand,
many of the nontrivial dynamical properties of the Nambu--Jona-Lasinio
model remain valid in two dimensions and thus allow its study in a 
consistent way in a simpler framework. This model was considered by 
Gross and Neveu \cite{Gross:1974jv} and is called after them.
\par  
In addition to the global $U(N)$ symmetry, the system satisfies 
also a discrete chiral symmetry, which prevents the fermions
from having a mass. In more general versions of the model, 
a continuous chiral symmetry is imposed rather than a discrete one,
but the latter is already sufficient to reproduce the nontrivial
effects of the dynamics. We stick here to the discrete chiral
symmetry.
\par
The Lagrangian density of the system is   
\be \lb{e39}
\mathcal{L}=\overline\psi^ai\gamma^{\mu}\partial_{\mu}\psi^a
+\frac{g_0}{2N}(\overline\psi^a\psi^a)^2,\ \ \ \ \ g_0>0. 
\ee
(The positivity of $g_0$ is justified in Ref. \cite{Gross:1974jv}.)
Summation on repeated indices is understood; $a$ runs from $1$ to $N$.
$\overline\psi^a=\psi^{\dagger a}\gamma^0$.
In two space-time dimensions, the Dirac fields have two components 
and the Dirac matrices $\gamma$ reduce to the Pauli
$2\times 2$ matrices $\sigma$:
\be \lb{e40} 
\gamma^0=\sigma_z,\ \ \ \ \gamma^1=i\sigma_y,\ \ \ \ 
\gamma_5=\gamma^0\gamma^1=\sigma_x.
\ee
The fermion fields are two-component spinors, with indices
$\alpha$ or $\beta$ ($\alpha,\beta=1,2$).   
Spinor indices will often be omitted from our notations; in some
cases, for instance in the Lagrangian density or in mass terms,
there is an implicit summation on them together with those of the
$\gamma$ matrices.
\par
Discrete chiral transformations are defined in the following way:
\be \lb{e41}
\psi^a\rightarrow \gamma_5\psi^a,\ \ \ \ \ 
\overline\psi^a=\psi^{\dagger a}\gamma^0\rightarrow -\overline\psi^a\gamma_5.
\ee
A mass term is not invariant under these transformations:
\be \lb{e42} 
m\overline\psi^a\psi^a\rightarrow -m\overline\psi^a\psi^a.
\ee
Hence, the fermions should be massless (no bare mass term in the 
Lagrangian density) if discrete chiral invariance is imposed on 
the theory.
\par    
The fermion field has mass dimension 1/2. As a consequence, the 
bare coupling constant $g_0$ is dimensionless.
\par 

\subsection{Asymptotic freedom} \lb{s32}

The analysis of the theory is done in much the same way as for
the $O(N)$ model. We first introduce the auxiliary field $\sigma$
and then retain the leading terms in the $1/N$ expansion.
\par
The auxiliary field is introduced with the addition of a new
term to the Lagrangian density:
\be \lb{e43} 
\mathcal{L'}=\mathcal{L}-\frac{N}{2g_0}
\Big(\sigma+\frac{g_0}{N}\ \overline\psi^a\psi^a\Big)^2.
\ee
The field $\sigma$ is defined through its equation of motion:
\be \lb{e43b}
\sigma+\frac{g_0}{N}\ \overline\psi^a\psi^a=0.
\ee
$\mathcal{L'}$ takes the form
\be \lb{e44}
\mathcal{L'}=\overline\psi^ai\gamma^{\mu}\partial_{\mu}\psi^a
-\frac{N}{2g_0}\sigma^2-\sigma\ \overline\psi^a\psi^a.
\ee 
\par
The bare fermion propagator is defined as
\be \lb{e45}
S_{0,\alpha\beta}^{ab}(p)=\delta_{ab}S_{0,\alpha\beta}(p)=
\int d^4x e^{{\displaystyle ip.x}}
<0|T(\psi_{\alpha}^a(x)\overline{\psi}_{\beta}^b(0))|0>,
\ee
where $\alpha$ and $\beta$ are the fermion field spinor indices
and $S_0$ is given by the expression 
\be \lb{e46}
S_{0}(p)=\frac{i\gamma.p}{p^2+i\varepsilon}\ \ \Longleftrightarrow \ \ 
\input{f3-nexpm.pstex_t}\ \ \ = O(N^0).
\ee
\par
The bare $\sigma$ propagator is
\be \lb{e47}
D_{0\sigma}(p)=-\frac{ig_0}{N}\ \ \Longleftrightarrow \ \ 
\input{f7-nexpm.pstex_t}\ \ \ = O(N^{-1}).
\ee
\par
The bare vertex $\sigma\overline{\psi}^a\psi^a$, with coefficient $-i$, 
has a graphical representation similar to that of Fig. \rf{f6}.
\par
The orders in $N$, when $N$ is large, of the propagators, the vertex and 
the loops are the same as those found in the $O(N)$ model, the $\psi$s
replacing now the $\phi$s. The leading part in the loop contributions
to the effective potential will come from single fermion loops associated 
with constant external sigma fields. 
\par
In the calculation of the effective potential one considers constant
classical external fields. Such fields may be interpreted as the
vacuum expectation value of the quantized field operator:
$\phi_{\mathrm{class}}=<0|\phi(x)|0>$. Because of the translation 
invariance of the vacuum state, the vacuum expectation value of the 
field operator is $x$ independent and hence $\phi_{\mathrm{class}}$
is constant. This property can be applied to scalar fields. When
the field is a fermion, it transforms under Lorentz
transformations as a spinor, while the vaccum state remains invariant.
This immediately implies that the vacuum expectation value of a 
fermion field is zero. Therefore, for the calculation of the effective
potential, one has to consider only external scalar fields (in our
case, the $\sigma$ field). Fermions contribute only in internal
lines or loops. 
\par
Another property of fermion fields appears also in loop calculations.
Loops of fermion fields involve at the end the trace operation on the
spinor indices. For massless fermions, the propagator is proportional 
to the  matrix $\gamma.p$. The loop value of a single fermion (without
any external line) is proportional to the trace of the $\gamma$ matrix,
which is zero. This property generalizes easily to loops associated with
an odd number of external constant scalar fields. Therefore, the
quantum part of the effective potential will involve only fermion loops 
with an even number of external lines.
\par
The effective potential is given as usual by the sum of two 
contributions [Eq. (\rf{e16})]:
\be \lb{e48}
U_{\mathrm{eff}}=U_{\mathrm{class}}+U_{\mathrm{rad}}\ ,
\ee
where $U_{\mathrm{class}}=\frac{N}{2g_0}\sigma^2$.
$U_{\mathrm{eff}}$ is calculated by the infinite sum of the 
diagrams of Fig. \rf{f10}.
\par
\bfg 
\bc
\input{f11-nexpm.pstex_t}
\caption{Diagrams, at leading order in the $1/N$ expansion, composing
$U_{\mathrm{eff}}$. The first diagram represents $U_{\mathrm{class}}$.}
\lb{f10}
\ec
\efg
One obtains:
\be \lb{e49}
U_{\mathrm{eff}}=N\Big(\frac{\sigma^2}{2g_0}-\int \frac{d^2k}{(2\pi)^2}
\ln(1+\frac{\sigma^2}{k^2})\Big).
\ee
\be \lb{e50}
\frac{\partial U_{\mathrm{eff}}}{\partial \sigma}=
2\sigma N\Big(\frac{1}{2g_0}
-\int \frac{d^2k}{(2\pi)^2}\ \frac{1}{(k^2+\sigma^2)}\Big).
\ee
(The integration momenta are euclidean; see comment before Eq. (\rf{e18}).)
Concentrating on Eq. (\rf{e50}), we have to isolate, as for the scalar 
case, the divergent part of the integral. We introduce a mass parameter 
$\mu$ in the subtracted term to avoid infrared divergence:
\be \lb{e51}
\frac{\partial U_{\mathrm{eff}}}{\partial \sigma}=
\sigma N\Big[\frac{1}{g_0}
-2\int \frac{d^2k}{(2\pi)^2}\ \frac{1}{(k^2+\mu^2)}\Big]
+\frac{\sigma N}{2\pi}\ln(\frac{\sigma^2}{\mu^2}).
\ee
We deduce from the latter expression the renormalization of the coupling 
constant into a finite value:
\be \lb{e52}
\frac{1}{g(\mu)}=\frac{1}{g_0}-2\int \frac{d^2k}{(2\pi)^2}\ 
\frac{1}{(k^2+\mu^2)},\ \ \ \ g>0.
\ee
\par
$g$ depends on the mass parameter $\mu$.
Its relation to another choice, $\mu_1$, with
$g(\mu_1)\equiv g_1$, is:
\be \lb{e53}
\frac{1}{g(\mu)}=\frac{1}{g_1}+\frac{1}{2\pi}\ln(\frac{\mu^2}{\mu_1^2}),
\ee
or,
\be \lb{e54}
g(\mu)=\frac{g_1}{1+\frac{g_1}{2\pi}\ln(\frac{\mu^2}{\mu_1^2})}.
\ee
\par
$g(\mu)$ decreases when $\mu$ increases. We also have:
\be \lb{e55}
\mu\frac{\partial g}{\partial \mu}=-\frac{g^2}{\pi}=\beta(g)<0.
\ee
The theory is therefore asymptotically free. This means that at
high energies or at short distances, the interaction becomes
weaker and weaker. In these regions perturbation theory can be
used with respect to the coupling constant $g(\mu)$. On the 
opposite side, at low energies or at large distances, the
interaction becomes strong. There is even a critical value of 
$\mu$, $\mu_{\mathrm{crit}}=\mu_1\exp{(-\pi/g_1)}$, for which
$g(\mu)$ diverges, indicating the occurrence of instabilities 
in the theory.
\par

\subsection{The ground state} \lb{3s3}

The occurrence of instabilities may be rather a sign that the
ground state that we are considering and around which 
calculations are done with free field propagators, is not the
correct one. We have to determine from the effective potential
the true ground state of the theory.
\par  
With respect to the renormalized coupling constant $g(\mu)$,
the effective potential takes the form
\be \lb{e56}
U_{\mathrm{eff}}=\frac{N\sigma^2}{4\pi}\ \Big[\ \frac{2\pi}{g(\mu)}
-1+\ln\big(\frac{\sigma^2}{\mu^2}\big)\ \Big].
\ee
The minimum of the effective potential is obtained from the
equation
\be \lb{e57}
\frac{\partial U_{\mathrm{eff}}}{\partial \sigma}=
\frac{N\sigma}{2\pi}\ \Big[\ \frac{2\pi}{g(\mu)}
+\ln(\frac{\sigma^2}{\mu^2})\ \Big]=0,
\ee
which possesses two types of solution: 1) $\sigma=0$; 
2) $\sigma=\pm\mu e^{-\pi/g(\mu)}$.
\par
The absolute minimum of $U_{\mathrm{eff}}$ can be searched for
with a study of the shape of the function $U_{\mathrm{eff}}$. 
The latter is represented graphically in Fig. \rf{f11}.
\par
\bfg
\bc
\epsfig{file=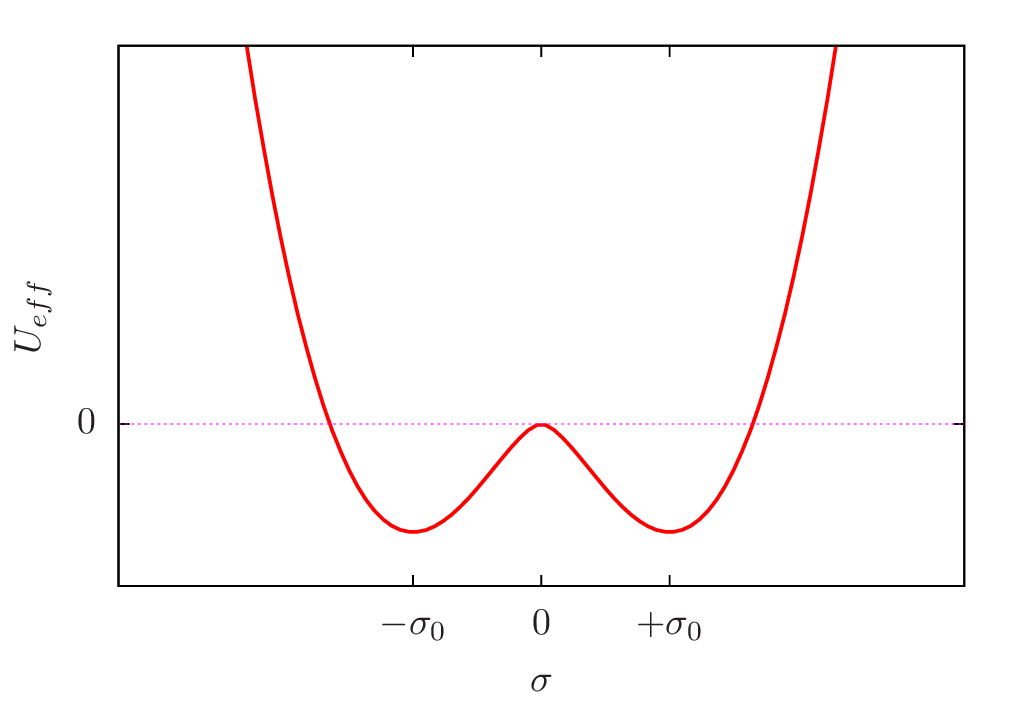,scale=1.2}
\caption{The shape of the function $U_{\mathrm{eff}}$ with respect 
to $\sigma$. We have defined $\sigma_0=\mu e^{-\pi/g(\mu)}$.}
\lb{f11}
\ec\efg
One notes that the solution $\sigma=0$ is a local maximum, while the
solutions $\sigma=\pm\mu e^{-\pi/g(\mu)}$ represent degenerate absolute 
minima. Because of the symmetry of $U_{\mathrm{eff}}$ under the change
of sign of $\sigma$, any of these can be equivalently chosen. We shall
choose for definiteness the solution with a plus sign and shall
designate it by $\sigma_0$:
\be \lb{e58}
\sigma_0=\mu e^{{\displaystyle -\pi/g(\mu)}}.
\ee
In order to study the physical properties of the system around the
new ground state, we must shift the field $\sigma$ in the effective 
potential and in the Lagrangian density by $\sigma_0$:
\be \lb{e59}
\sigma=\sigma'+\sigma_0.
\ee
\par
In terms of the new field $\sigma'$, the effective potential 
becomes
\be \lb{e60}
U_{\mathrm{eff}}=\frac{N\sigma_0^2}{4\pi}\ 
\big(1+\frac{\sigma'}{\sigma_0}\big)^2\
\Big[\ -1+\ln\big(1+\frac{\sigma'}{\sigma_0}\big)^2\ \Big].
\ee
We find that $g(\mu)$ and $\mu$ have completely disappeared from the 
new expression of $U_{\mathrm{eff}}$  in favor of the single
parameter $\sigma_0$, which has, by the definition of $\sigma$
[Eq. (\rf{e43b})], a dimension of mass and thus fixes the mass scale
of the theory. No free adjustable parameter has remained.
\par
The Lagrangian density becomes:
\be \lb{e61}
\mathcal{L'}=\overline\psi^ai\gamma^{\mu}\partial_{\mu}\psi^a
-\frac{N}{2g_0}(\sigma'+\sigma_0)^2-\sigma_0\overline\psi^a\psi^a
-\sigma'\overline\psi^a\psi^a.
\ee 
We note the appearance of a mass term of fermions, with value
\be \lb{e62}
M_F=\sigma_0=\mu e^{{\displaystyle -\pi/g(\mu)}}.
\ee
The Lagrangian density still contains the bare coupling constant
$g_0$. This is necessary, since one still has to calculate and
renormalize the $\sigma'$ propagator. The latter receives
contributions from the fermion loops which are divergent and whose
divergence should be combined with $g_0$ to reproduce a finite quantity.
\par
The theory was invariant at the beginning under the discrete chiral
transformations, imposing masslessness of the fermions, but now,
after renormalization and the shift to the ground state of the
energy, the fermions have acquired a mass. This phenomenon is called
dynamical mass generation and is due to the spontaneous 
breaking of the discrete chiral symmetry. 
\par
On the other hand, $M_F$ is a physical quantity and
should not depend on the particular choices of the arbitrary
mass parameter $\mu$. From Eq. (\rf{e62}) one notes that $M_F$ has
two types of dependence on $\mu$: an explicit one and an implicit one
through the coupling constant $g(\mu)$ [Eq. (\rf{e55})]. One verifies 
that $M_F$ is actually independent of $\mu$:
\be \lb{e63}
\frac{d M_F}{d\mu}=\frac{\partial M_F}{\partial \mu}
+\frac{1}{\mu}\beta(g)\frac{\partial M_F}{\partial g}=0.
\ee
$M_F$ is said to be renormalization group invariant.
\par
Once $M_F$ is fixed, the two other parameters, $g$ and
$\mu$ should disappear from the finite quantities of the theory. 
Any choice of $\mu$ is compensated by a corresponding choice of 
$g$, to produce $M_F$. This property was explicitly verified
on the expression of $U_{\mathrm{eff}}$ [Eq. (\rf{e60})].
\par
At the beginning, the theory had a dimensionless parameter 
$g$ and massless fermions. Now it has instead a dimensionful parameter, 
$M_F$, mass of the fermions, which is fixed by the 
physical conditions and sets the mass scale of the theory. 
There is no longer a free parameter in the theory. This phenomenon 
is called mass transmutation.
\par   
In summary, the Gross-Neveu model displays many interesting features of
quantum field theory and illustrates, in two dimensions, several
phenomena -- asymptotic freedom, dynamical mass generation,
mass transmutation -- expected also to occur in four-dimensional
theories.
\par

\section{QCD} \lb{s4}

Quantum Chromodynamics (QCD), the theory of the strong interaction, 
is a gauge theory with the non-Abelian local symmetry group of color 
$SU(N)_c$, with $N=3$. We consider henceforth the general case where 
the paramater $N$ is arbitrary. As ``matter'' fields, the
theory contains quark and antiquark fields, belonging to the defining 
fundamental representation of the group and to its conjugate, respectively; 
their number is $N$. Since this is a gauge theory, there are also gauge 
fields, the gluons, belonging to the adjoint representation of the group; 
their number is $(N^2-1)$.
\par
There are six ``generations'' of QCD, distinguished from each
other by specific properties of the quarks (charge and mass), called
also ``flavors''. Schematically, there are three ``light'' quarks,
$u$, $d$, $s$ and three ``heavy'' quarks, $c$, $b$, $t$. The free
masses of the quarks $u$ and $d$ are of the order of a few MeV, while
the mass of the quark $s$ is of the order of 100 MeV. The masses
of the heavy quarks $c$, $b$, $t$ are approximately 1.3 GeV, 4.2 GeV
and 173 GeV, respectively \cite{Agashe:2014kda}. The gluon, which
is the gauge particle, is massless.  
\par
The proton, the stable matter particle, is a bound state made mainly
of the three quarks $u$, $u$, $d$. Its mass is of the order of 1 GeV. 
This shows that the mass of the proton is not made of the masses of its
quark components, which are nearly massless. Rather, one should expect 
that it is produced by a dynamical mass generation mechanism, similar
to what happened in the Gross-Neveu model. A similar conclusion also
holds for the other low-lying hadrons (neutron, $\rho$ meson, etc.).
On the other hand, the coupling constant of the QCD Lagrangian is 
dimensionless. This means that the mass generation phenomenon would
be realized by the mechanism of mass transmutation, which also was 
observed in the Gross-Neveu model.
\par
Since the light quark masses do not seem to play a fundamental role,
one can consider the QCD Lagrangian in the ideal situation where the
three light quarks are massless. In this case, the QCD theory
also satisfies a global flavor space invariance under the group of 
continuous chiral transformations $SU(3)_R\times SU(3)_L$ ($R$ for right, 
$L$ for left). It is expected that this symmetry is realized with
the Goldstone mode (no nearly degenerate parity doublets are observed 
in nature); the corresponding Goldstone bosons are the lowest lying
$\pi$, $K$ and $\eta$ mesons. In the limit of vanishing quark masses,
the masses of the latter particles would also vanish. For small values 
of the quark masses, the Goldstone bosons also acquire a small mass.
This explains why these mesons have masses squared much smaller than
the other hadron masses squared.
\par
QCD theory has been widely investigated in ordinary perturbation
theory and has been shown to be asymptotically free 
\cite{Gross:1973id,Politzer:1973fx}; a relation similar to Eq. 
(\rf{e54}) has been obtained. This implies that at high energies
or at short distances, the QCD interaction becomes weak and 
ordinary perturbation theory can be applied there. This property
has been experimentally verified by many high-energy experiments.
The other implication of asymptotic freedom is that at low energies
or at large distances the interaction becomes strong enough to 
forbid perturbative treatments. 
From asymptotic freedom one also deduces the existence of a
renormalization group invariant mass, called $\Lambda_{\mathrm{QCD}}$,
which realizes mass transmutation in the theory (cf. Eqs. (\rf{e58}),
(\rf{e62}), (\rf{e63})). Contrary to the Gross-Neveu model, however,
the resolution of the nonperturbative domain of QCD has not been 
achieved up to now with analytic calculations.
One of the main reasons of this failure is probably related 
to the fact that quarks and gluons are confined. These particles
have not been observed as free asymptotic states like the other
known particles. Their presence or existence have been detected 
mainly in an indirect way. Quarks and gluons are bound by the QCD
force to form bound states called hadrons (proton, neutron, $\pi$ and
$\rho$ mesons, etc.) It is at this level that QCD differs from the
previous models that were considered or mentioned (Gross-Neveu, 
Nambu--Jona-Lasinio). Numerical resolution of the strong coupling
regime of QCD is successfully realized with Lattice calculations.
\par   
The application of the $1/N$ expansion method to QCD leads to 
some simplifications and well verified qualitative predictions, 
but fails to solve the theory as a whole. The reason of this last 
negative result is due to the large number of gluon fields
($\sim N^2$) as compared to that of the quarks ($N$). While quark loop
contributions become negiligible and do not enter in leading 
expressions in the large-$N$ limit, gluon loop diagrams, and more
precisely the class of ``planar diagrams'', become dominant at 
large $N$ and an infinite number of them (one-particle irreducible) 
survive \cite{'tHooft:1973jz,Witten:1979kh,Witten:1979pi,Coleman:1980nk,
Makeenko:1999hq}. 
Their summation in compact form is not an easy task.
Examples of planar and non-planar diagrams are
presented in Figs. \rf{f12} and \rf{f13}, respectively.
\par
\bfg
\input{f14-nexpm.tex}
\caption{Examples of planar diagrams. Curly lines represent gluons
and oriented full lines quarks or antiquarks. When a planar diagram
is drawn on a plane, a gluon line does not intersect any other gluon
line, except at vertices. The latter are represented on the figures
by dots.}
\lb{f12}
\efg 
\par
\bfg
\input{f15-nexpm.tex}
\caption{Examples of non-planer or crossed diagrams. The intersection
points of the gluon lines are not vertices.}
\lb{f13}
\efg
\par
One therefore is satisfied at the present time with the qualitative
predictions and relative simplifications the $1/N$ expansion method
provides. The method, when applied to two-dimensional QCD, has been,
however, successful enough to solve the main aspects of the theory in 
an explicit way \cite{'tHooft:1974hx,Callan:1975ps}. 
\par

\section{Conclusion} \lb{s5}

The $1/N$ expansion method allows us to solve, at
leading order of the expansion, theories and models nonperturbatively,
probing directly dynamical phenomena, which would not be reached in
ordinary perturbation theory based on expansions with respect to 
the coupling constant.
\par
The method is also applicable to QCD, but because of the presence
of the gluon fields, which belong to the adjoint representation of the
gauge symmetry group, its effects are less spectacular. Nevertheless,
many simplifications occur and several qualitative features can be
drawn about the properties of the theory.
\par

\vspace{0.5 cm}
\noindent
\textbf{Acknowledgements}
\par
I thank the Organizing Committee and its members for their invitation 
and kind hospitality and for the stimulating atmosphere created at the 
Summer School.
This work was partially supported by the EU I3HP Project ''Study of
Strongly Interacting Matter'' (acronym HadronPhysics3, Grant
Agreement No. 283286).
\par

\end{document}

%% file: f3-nexpm.pstex_t
\begin{picture}(0,0)%
\includegraphics{f3-nexpm.pstex}%
\end{picture}%
\setlength{\unitlength}{3108sp}%
\begingroup\makeatletter\ifx\SetFigFont\undefined%
\gdef\SetFigFont#1#2#3#4#5{%
  \reset@font\fontsize{#1}{#2pt}%
  \fontfamily{#3}\fontseries{#4}\fontshape{#5}%
  \selectfont}%
\fi\endgroup%
\begin{picture}(2091,66)(2893,-4369)
\end{picture}%

%% file: f4-nexpm.pstex_t
\begin{picture}(0,0)%
\includegraphics{f4-nexpm.pstex}%
\end{picture}%
\setlength{\unitlength}{3108sp}%
\begingroup\makeatletter\ifx\SetFigFont\undefined%
\gdef\SetFigFont#1#2#3#4#5{%
  \reset@font\fontsize{#1}{#2pt}%
  \fontfamily{#3}\fontseries{#4}\fontshape{#5}%
  \selectfont}%
\fi\endgroup%
\begin{picture}(1956,2314)(3073,-5174)
\put(3331,-3031){\makebox(0,0)[lb]{\smash{{\SetFigFont{11}{13.2}{\familydefault}{\mddefault}{\updefault}{\color[rgb]{0,0,0}$a$}%
}}}}
\put(3376,-5101){\makebox(0,0)[lb]{\smash{{\SetFigFont{11}{13.2}{\familydefault}{\mddefault}{\updefault}{\color[rgb]{0,0,0}$a$}%
}}}}
\put(4501,-3076){\makebox(0,0)[lb]{\smash{{\SetFigFont{11}{13.2}{\familydefault}{\mddefault}{\updefault}{\color[rgb]{0,0,0}$b$}%
}}}}
\put(4546,-5101){\makebox(0,0)[lb]{\smash{{\SetFigFont{11}{13.2}{\familydefault}{\mddefault}{\updefault}{\color[rgb]{0,0,0}$b$}%
}}}}
\put(4321,-4066){\makebox(0,0)[lb]{\smash{{\SetFigFont{11}{13.2}{\familydefault}{\mddefault}{\updefault}{\color[rgb]{0,0,0}$-i\frac{\lambda_0}{8N}=O(N^{-1})$}%
}}}}
\end{picture}%

%% file: f5-nexpm.pstex_t
\begin{picture}(0,0)%
\includegraphics{f5-nexpm.pstex}%
\end{picture}%
\setlength{\unitlength}{2486sp}%
\begingroup\makeatletter\ifx\SetFigFont\undefined%
\gdef\SetFigFont#1#2#3#4#5{%
  \reset@font\fontsize{#1}{#2pt}%
  \fontfamily{#3}\fontseries{#4}\fontshape{#5}%
  \selectfont}%
\fi\endgroup%
\begin{picture}(10866,2638)(868,-5363)
\put(946,-3526){\makebox(0,0)[lb]{\smash{{\SetFigFont{9}{10.8}{\familydefault}{\mddefault}{\updefault}{\color[rgb]{0,0,0}$a$}%
}}}}
\put(3151,-3571){\makebox(0,0)[lb]{\smash{{\SetFigFont{9}{10.8}{\familydefault}{\mddefault}{\updefault}{\color[rgb]{0,0,0}$a$}%
}}}}
\put(2026,-2896){\makebox(0,0)[lb]{\smash{{\SetFigFont{9}{10.8}{\familydefault}{\mddefault}{\updefault}{\color[rgb]{0,0,0}$b$}%
}}}}
\put(4951,-2896){\makebox(0,0)[lb]{\smash{{\SetFigFont{9}{10.8}{\familydefault}{\mddefault}{\updefault}{\color[rgb]{0,0,0}$a$}%
}}}}
\put(4861,-4741){\makebox(0,0)[lb]{\smash{{\SetFigFont{9}{10.8}{\familydefault}{\mddefault}{\updefault}{\color[rgb]{0,0,0}$a$}%
}}}}
\put(7201,-2941){\makebox(0,0)[lb]{\smash{{\SetFigFont{9}{10.8}{\familydefault}{\mddefault}{\updefault}{\color[rgb]{0,0,0}$b$}%
}}}}
\put(7156,-4651){\makebox(0,0)[lb]{\smash{{\SetFigFont{9}{10.8}{\familydefault}{\mddefault}{\updefault}{\color[rgb]{0,0,0}$b$}%
}}}}
\put(6121,-4516){\makebox(0,0)[lb]{\smash{{\SetFigFont{9}{10.8}{\familydefault}{\mddefault}{\updefault}{\color[rgb]{0,0,0}$c$}%
}}}}
\put(8866,-3031){\makebox(0,0)[lb]{\smash{{\SetFigFont{9}{10.8}{\familydefault}{\mddefault}{\updefault}{\color[rgb]{0,0,0}$a$}%
}}}}
\put(9991,-3256){\makebox(0,0)[lb]{\smash{{\SetFigFont{9}{10.8}{\familydefault}{\mddefault}{\updefault}{\color[rgb]{0,0,0}$a$}%
}}}}
\put(11251,-3121){\makebox(0,0)[lb]{\smash{{\SetFigFont{9}{10.8}{\familydefault}{\mddefault}{\updefault}{\color[rgb]{0,0,0}$a$}%
}}}}
\put(8821,-4876){\makebox(0,0)[lb]{\smash{{\SetFigFont{9}{10.8}{\familydefault}{\mddefault}{\updefault}{\color[rgb]{0,0,0}$b$}%
}}}}
\put(10036,-4696){\makebox(0,0)[lb]{\smash{{\SetFigFont{9}{10.8}{\familydefault}{\mddefault}{\updefault}{\color[rgb]{0,0,0}$b$}%
}}}}
\put(11206,-4831){\makebox(0,0)[lb]{\smash{{\SetFigFont{9}{10.8}{\familydefault}{\mddefault}{\updefault}{\color[rgb]{0,0,0}$b$}%
}}}}
\put(6031,-3121){\makebox(0,0)[lb]{\smash{{\SetFigFont{9}{10.8}{\familydefault}{\mddefault}{\updefault}{\color[rgb]{0,0,0}$c$}%
}}}}
\put(1846,-4336){\makebox(0,0)[lb]{\smash{{\SetFigFont{9}{10.8}{\familydefault}{\mddefault}{\updefault}{\color[rgb]{0,0,0}$O(N^0)$}%
}}}}
\put(5806,-5236){\makebox(0,0)[lb]{\smash{{\SetFigFont{9}{10.8}{\familydefault}{\mddefault}{\updefault}{\color[rgb]{0,0,0}$O(N^{-1})$}%
}}}}
\put(9856,-5281){\makebox(0,0)[lb]{\smash{{\SetFigFont{9}{10.8}{\familydefault}{\mddefault}{\updefault}{\color[rgb]{0,0,0}$O(N^{-2})$}%
}}}}
\end{picture}%

%% file: f6-nexpm.pstex_t
\begin{picture}(0,0)%
\includegraphics{f6-nexpm.pstex}%
\end{picture}%
\setlength{\unitlength}{2486sp}%
\begingroup\makeatletter\ifx\SetFigFont\undefined%
\gdef\SetFigFont#1#2#3#4#5{%
  \reset@font\fontsize{#1}{#2pt}%
  \fontfamily{#3}\fontseries{#4}\fontshape{#5}%
  \selectfont}%
\fi\endgroup%
\begin{picture}(8931,8201)(1093,-8329)
\end{picture}%

%% file: f7-nexpm.pstex_t
\begin{picture}(0,0)%
\includegraphics{f7-nexpm.pstex}%
\end{picture}%
\setlength{\unitlength}{3108sp}%
\begingroup\makeatletter\ifx\SetFigFont\undefined%
\gdef\SetFigFont#1#2#3#4#5{%
  \reset@font\fontsize{#1}{#2pt}%
  \fontfamily{#3}\fontseries{#4}\fontshape{#5}%
  \selectfont}%
\fi\endgroup%
\begin{picture}(2406,66)(1588,-3739)
\end{picture}%

%% file: f8-nexpm.pstex_t
\begin{picture}(0,0)%
\includegraphics{f8-nexpm.pstex}%
\end{picture}%
\setlength{\unitlength}{3108sp}%
\begingroup\makeatletter\ifx\SetFigFont\undefined%
\gdef\SetFigFont#1#2#3#4#5{%
  \reset@font\fontsize{#1}{#2pt}%
  \fontfamily{#3}\fontseries{#4}\fontshape{#5}%
  \selectfont}%
\fi\endgroup%
\begin{picture}(3243,1999)(2443,-4904)
\put(4636,-3076){\makebox(0,0)[lb]{\smash{{\SetFigFont{11}{13.2}{\familydefault}{\mddefault}{\updefault}{\color[rgb]{0,0,0}$a$}%
}}}}
\put(4681,-4831){\makebox(0,0)[lb]{\smash{{\SetFigFont{11}{13.2}{\familydefault}{\mddefault}{\updefault}{\color[rgb]{0,0,0}$a$}%
}}}}
\put(5671,-3976){\makebox(0,0)[lb]{\smash{{\SetFigFont{11}{13.2}{\familydefault}{\mddefault}{\updefault}{\color[rgb]{0,0,0}$=\ O(N^0)$}%
}}}}
\end{picture}%

%% file: f9-nexpm.pstex_t
\begin{picture}(0,0)%
\includegraphics{f9-nexpm.pstex}%
\end{picture}%
\setlength{\unitlength}{2486sp}%
\begingroup\makeatletter\ifx\SetFigFont\undefined%
\gdef\SetFigFont#1#2#3#4#5{%
  \reset@font\fontsize{#1}{#2pt}%
  \fontfamily{#3}\fontseries{#4}\fontshape{#5}%
  \selectfont}%
\fi\endgroup%
\begin{picture}(11226,6454)(868,-7388)
\put(10261,-7306){\makebox(0,0)[lb]{\smash{{\SetFigFont{9}{10.8}{\familydefault}{\mddefault}{\updefault}{\color[rgb]{0,0,0}$O(N^{-2})$}%
}}}}
\put(6166,-7306){\makebox(0,0)[lb]{\smash{{\SetFigFont{9}{10.8}{\familydefault}{\mddefault}{\updefault}{\color[rgb]{0,0,0}$O(N^{-1})$}%
}}}}
\put(2116,-7306){\makebox(0,0)[lb]{\smash{{\SetFigFont{9}{10.8}{\familydefault}{\mddefault}{\updefault}{\color[rgb]{0,0,0}  $O(N)$}%
}}}}
\put(4231,-3571){\makebox(0,0)[lb]{\smash{{\SetFigFont{9}{10.8}{\familydefault}{\mddefault}{\updefault}{\color[rgb]{0,0,0}$O(N^0)$}%
}}}}
\put(8146,-3571){\makebox(0,0)[lb]{\smash{{\SetFigFont{9}{10.8}{\familydefault}{\mddefault}{\updefault}{\color[rgb]{0,0,0}$O(N^0)$}%
}}}}
\end{picture}%

%% file: f12-nexpm.pstex_t
\begin{picture}(0,0)%
\includegraphics{f12-nexpm.pstex}%
\end{picture}%
\setlength{\unitlength}{2901sp}%
\begingroup\makeatletter\ifx\SetFigFont\undefined%
\gdef\SetFigFont#1#2#3#4#5{%
  \reset@font\fontsize{#1}{#2pt}%
  \fontfamily{#3}\fontseries{#4}\fontshape{#5}%
  \selectfont}%
\fi\endgroup%
\begin{picture}(7626,4359)(1003,-5678)
\put(2161,-5551){\makebox(0,0)[lb]{\smash{{\SetFigFont{11}{13.2}{\familydefault}{\mddefault}{\updefault}{\color[rgb]{0,0,0}$O(N^0)$}%
}}}}
\put(6616,-5596){\makebox(0,0)[lb]{\smash{{\SetFigFont{11}{13.2}{\familydefault}{\mddefault}{\updefault}{\color[rgb]{0,0,0}$O(N^{-1})$}%
}}}}
\put(4231,-2941){\makebox(0,0)[lb]{\smash{{\SetFigFont{11}{13.2}{\familydefault}{\mddefault}{\updefault}{\color[rgb]{0,0,0}$O(N)$}%
}}}}
\end{picture}%

%% file: f10-nexpm.pstex_t
\begin{picture}(0,0)%
\includegraphics{f10-nexpm.pstex}%
\end{picture}%
\setlength{\unitlength}{2486sp}%
\begingroup\makeatletter\ifx\SetFigFont\undefined%
\gdef\SetFigFont#1#2#3#4#5{%
  \reset@font\fontsize{#1}{#2pt}%
  \fontfamily{#3}\fontseries{#4}\fontshape{#5}%
  \selectfont}%
\fi\endgroup%
\begin{picture}(8815,3171)(1191,-5314)
\put(2566,-3841){\makebox(0,0)[lb]{\smash{{\SetFigFont{9}{10.8}{\familydefault}{\mddefault}{\updefault}{\color[rgb]{0,0,0}$+$}%
}}}}
\put(4816,-3841){\makebox(0,0)[lb]{\smash{{\SetFigFont{9}{10.8}{\familydefault}{\mddefault}{\updefault}{\color[rgb]{0,0,0}$+$}%
}}}}
\put(7381,-3886){\makebox(0,0)[lb]{\smash{{\SetFigFont{9}{10.8}{\familydefault}{\mddefault}{\updefault}{\color[rgb]{0,0,0}$+$}%
}}}}
\put(9991,-3841){\makebox(0,0)[lb]{\smash{{\SetFigFont{9}{10.8}{\familydefault}{\mddefault}{\updefault}{\color[rgb]{0,0,0}$+\ \cdots$}%
}}}}
\end{picture}%

%% file: f11-nexpm.pstex_t
\begin{picture}(0,0)%
\includegraphics{f11-nexpm.pstex}%
\end{picture}%
\setlength{\unitlength}{2279sp}%
\begingroup\makeatletter\ifx\SetFigFont\undefined%
\gdef\SetFigFont#1#2#3#4#5{%
  \reset@font\fontsize{#1}{#2pt}%
  \fontfamily{#3}\fontseries{#4}\fontshape{#5}%
  \selectfont}%
\fi\endgroup%
\begin{picture}(11118,3126)(1048,-4864)
\put(7921,-3346){\makebox(0,0)[lb]{\smash{{\SetFigFont{8}{9.6}{\familydefault}{\mddefault}{\updefault}{\color[rgb]{0,0,0}$+$}%
}}}}
\put(3466,-3301){\makebox(0,0)[lb]{\smash{{\SetFigFont{8}{9.6}{\familydefault}{\mddefault}{\updefault}{\color[rgb]{0,0,0}$+$}%
}}}}
\put(12151,-3301){\makebox(0,0)[lb]{\smash{{\SetFigFont{8}{9.6}{\familydefault}{\mddefault}{\updefault}{\color[rgb]{0,0,0}$+\ \ \cdots$}%
}}}}
\end{picture}%

%% file: f14-nexpm.tex
\begin{center}
\begin{picture}(400,100)(0,0) 
\SetScale{0.75}

\Gluon(10,50)(70,50){3}{4}
\Vertex(70,50){2}
\Gluon(150,50)(210,50){3}{4}
\Vertex(150,50){2}
\GlueArc(110,50)(40,0,60){3}{3}
\Vertex(130,84.64){2}
\GlueArc(110,50)(40,60,120){3}{3}
\Vertex(90,84.64){2}
\GlueArc(110,50)(40,120,180){3}{3}
\GlueArc(110,50)(40,180,240){3}{3}
\Vertex(90,15.36){2}
\GlueArc(110,50)(40,240,300){3}{3}
\Vertex(130,15.36){2}
\GlueArc(110,50)(40,300,360){3}{3}
\Gluon(90,84.64)(90,50){3}{3}
\Vertex(90,50){2}
\Gluon(90,50)(90,15.36){3}{3}
\Gluon(130,84.64)(130,50){3}{3}
\Vertex(130,50){2}
\Gluon(130,50)(130,15.36){3}{3}
\Gluon(90,50)(130,50){3}{3}

\ArrowLine(300,90)(350,90)
\Vertex(350,90){2}
\ArrowLine(350,90)(400,90)
\Vertex(400,90){2}
\ArrowLine(400,90)(450,90)
\Vertex(450,90){2}
\ArrowLine(450,90)(500,90)
\ArrowLine(500,10)(450,10)
\Vertex(450,10){2}
\ArrowLine(450,10)(400,10)
\Vertex(400,10){2}
\ArrowLine(400,10)(350,10)
\Vertex(350,10){2}
\ArrowLine(350,10)(300,10)
\Gluon(350,90)(350,50){3}{3}
\Vertex(350,50){2}
\Gluon(350,50)(350,10){3}{3}
\Gluon(400,90)(400,50){3}{3}
\Vertex(400,50){2}
\Gluon(400,50)(400,10){3}{3}
\Gluon(350,50)(400,50){3}{3}
\Gluon(450,90)(450,10){3}{6}

\end{picture}
\end{center}

%% file: f15-nexpm.tex
\begin{center}
\begin{picture}(400,100)(0,0) 
\SetScale{0.75}

\Gluon(10,50)(70,50){3}{4}
\Vertex(70,50){2}
\Gluon(150,50)(210,50){3}{4}
\Vertex(150,50){2}
\GlueArc(110,50)(40,0,60){3}{3}
\Vertex(130,84.64){2}
\GlueArc(110,50)(40,60,120){3}{3}
\Vertex(90,84.64){2}
\GlueArc(110,50)(40,120,180){3}{3}
\GlueArc(110,50)(40,180,240){3}{3}
\Vertex(90,15.36){2}
\GlueArc(110,50)(40,240,300){3}{3}
\Vertex(130,15.36){2}
\GlueArc(110,50)(40,300,360){3}{3}
\Gluon(90,84.64)(130,15.36){3}{6}
\Gluon(130,84.64)(90,15.36){3}{6}

\ArrowLine(300,90)(350,90)
\Vertex(350,90){2}
\ArrowLine(350,90)(400,90)
\Vertex(400,90){2}
\ArrowLine(400,90)(450,90)
\Vertex(450,90){2}
\ArrowLine(450,90)(500,90)
\ArrowLine(500,10)(450,10)
\Vertex(450,10){2}
\ArrowLine(450,10)(400,10)
\Vertex(400,10){2}
\ArrowLine(400,10)(350,10)
\Vertex(350,10){2}
\ArrowLine(350,10)(300,10)
\Gluon(350,90)(400,10){3}{6}
\Gluon(400,90)(350,10){3}{6}
\Gluon(450,90)(450,10){3}{6}

\end{picture}
\end{center}